\iftrue
\documentclass[aps,prl,twocolumn,
			   groupedaddress,superscriptaddress,
			   amsfonts,amssymb,amsmath,
			   citeautoscript,
			   a4paper]{revtex4-1}
\else
\documentclass[aps,pra,preprint,
			   groupedaddress,superscriptaddress,notitlepage,
			   amsfonts,amssymb,amsmath,
			   citeautoscript,
			   a4paper]{revtex4-1}
\fi
\usepackage{mathtools}  
\usepackage{lmodern}    
\newenvironment{acknowledgement}{%
  \section*{Acknowledgements}%
}{}
\usepackage[pdftex]{hyperref}
\hypersetup{colorlinks,
			linkcolor={blue!75!black!80!yellow},
			citecolor={blue!75!black!80!yellow},
			urlcolor={blue!75!black!80!yellow},
			pdfstartview=FitH}
\usepackage{graphicx}
\usepackage{mathrsfs}
\usepackage{amsthm}
\usepackage{physics}
\usepackage{xspace}
\usepackage{braket}
\usepackage{xr}
\usepackage{gensymb} 
\usepackage{xcolor,soul}
\usepackage{stmaryrd} 
\usepackage[UKenglish]{babel}
\usepackage{placeins} 
\usepackage{siunitx}
\DeclareSIUnit[number-unit-product=]\percent{\char`\%} 
\usepackage{makecell}
\usepackage{diagbox}
\usepackage{float}
\usepackage{amsmath} 
\usepackage{empheq} 

\usepackage{txfonts}  
\usepackage{txfontsb} 
\usepackage{xr-hyper}
\usepackage{hyperref}

\renewcommand{\Re}{\operatorname{Re}}
\renewcommand{\Im}{\operatorname{Im}}
\newcommand{\iu}{\mathrm{i}}

\newcommand{\pt}{\mathcal{PT}} 
\renewcommand*\Tr{\mathop{}\mathrm{Tr}}

\usepackage{accents}

\usepackage{textcomp} 
\usepackage{xifthen}
\usepackage{etoolbox}
\newboolean{togglecomments}
\newboolean{togglechanges} 

\setboolean{togglecomments}{true}
\setboolean{togglechanges}{false}

\newcommand{\comment}[2]{%
    \ifbool{togglecomments}%
    {\textcolor{blue!70!black}{\small\textsf{%
    \textsuperscript{\textsc{\textsf{\MakeLowercase{#1}}}}%
    [#2]}}} 
    {}}     
\newcommand{\swap}[2]{\ifbool{togglechanges}
    {#2}  
    {\textcolor{red!70!black}{[#1]}\textrightarrow{}\textcolor{green!50!black}{[#2]}}}
\newcommand{\remove}[1]{\ifbool{togglechanges}
    {}    
    {\textcolor{red!70!black}{#1}}}
\newcommand{\inset}[1]{\ifbool{togglechanges}
    {#1}  
    {\textcolor{green!50!black}{#1}}}
\newcommand{\optional}[1]{\ifbool{togglechanges}
    {}    
    {\textcolor{yellow!50!orange!80!gray}{#1}}}

\newcommand{\citeremind}[1]{%
    [\textcolor{blue!75!black!80!yellow}{
        $\blacksquare$%
	    \ifthenelse{\isempty{#1}}
	        {}
	        {\textsuperscript{\tiny\textsf{#1}}}%
	}]\xspace}
        
\newcommand{\hkuaffil}{\footnotesize Department of Physics and HK Institute of Quantum Science and Technology, The University of Hong Kong, Pokfulam, Hong Kong, China}

\begin{document}

\title{Parity-time symmetry phase transition in photonic time-modulated media}

\author{Rui-Chuan Zhang}
\affiliation{\hkuaffil}
\author{Shu~Yang}
\affiliation{\hkuaffil}
\author{Yixin Sha}
\affiliation{\hkuaffil}
\author{Zetao Xie}
\affiliation{\hkuaffil}
\author{Yi~Yang}
\email{yiyg@hku.hk}
\affiliation{\hkuaffil}

\begin{abstract}
Time modulation can cause gain and loss in photonic media, leading to complex modal behaviors and enhanced wave controllability in the non-Hermitian regime. 
Conversely, we reveal that Hermiticity and parity-time (\(\mathcal{PT}\))-symmetry phase transition are possible under the temporal $\mathcal{PT}$-symmetry in time-modulated photonic media.
We prove that, for a homogeneously modulated photonic medium with complex-valued modulation, temporal $\mathcal{PT}$-symmetry is a necessary but insufficient condition for obtaining a real eigenvalue spectrum, giving rise to $\pt$-symmetry phase transition. 
Specifically, the $\pt$ phase transition critically depends on the contrast between the modulation depth of the real and imaginary parts of permittivity when they are sinusoidally modulated with a $\pi/2$ phase difference.
We generalize the discretized temporal-interface transfer matrix method to a continuous differential operator framework, which facilitates the confirmation of the phase transition condition via Magnus expansion analysis. 
Full-wave simulations and analytical calculations jointly confirm the occurrence of $\pt$-transition by examining the scattering behavior of a propagating pulse in such a type of modulated medium.
The findings provide a temporal $\pt$-symmetric paradigm for controlling Hermiticity and non-Hermiticity in spatiotemporal photonic systems.
\end{abstract}

\maketitle

\begin{figure}[h!]
    \centering
    \includegraphics[width=1\linewidth]{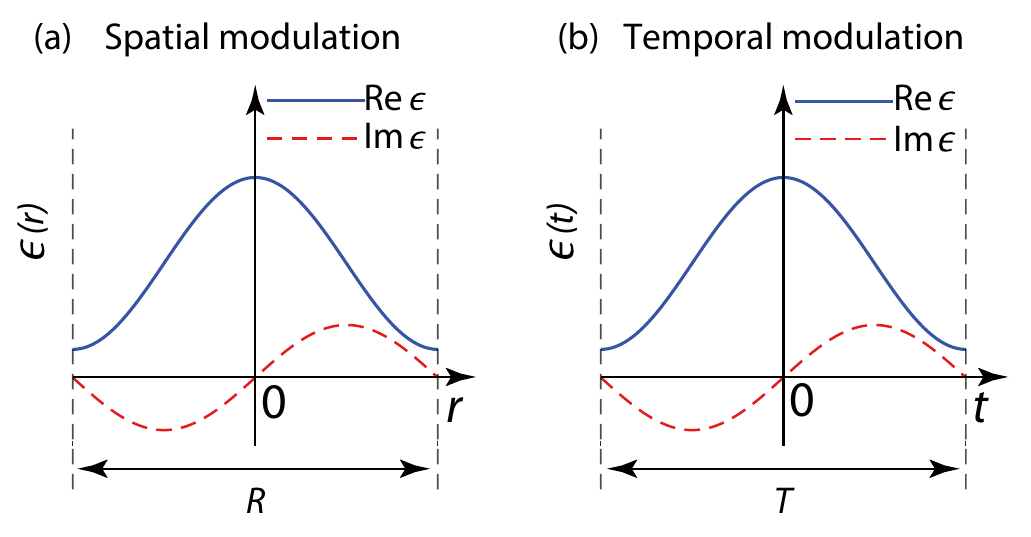} 
    \caption{
    \textbf{$\pt$-symmetry in spatial and temporal photonic crystals.}
     (a-b) Sinusoidal spatial (a) and temporal (b) modulation of permittivity that obeys $\pt$-symmetry. The spatial and temporal periodicity are noted as \(R\) and \(T\), respectively.
     }
    \label{fig:1}
\end{figure}

In Hermitian crystals with spatial periodicity, band gaps can emerge due to the periodic potential, ensuring well-defined real energy eigenstates~\cite{yablonovitch1987inhibited,john1987strong,joannopoulos2008molding}.
This condition can be further relaxed in non-Hermitian systems with the help of parity-time (\(\mathcal{PT}\))-symmetry~\cite{bender1998real, bender2007making}. 
In the past two decades, \(\mathcal{PT}\)-symmetry has been extensively studied in both quantum mechanics and photonics, providing a novel concept for exploring non-Hermitian phenomena~\cite{Regensburger2012, Feng2017, ozdemir2019parity,Alaeian2014, Cerjan2016, ozdemir2019parity, Mock2016,longhi2018parity,gupta2020parity,wang2023non}. 
In quantum mechanics, \(\mathcal{PT}\)-symmetry ensures that a non-Hermitian Hamiltonian can possess entirely real eigenvalues~\cite{bender1998real, bender2007making}. 
In photonics, this symmetry manifests through specific conditions on the permittivity $\epsilon(r)$ provided the system maintains a balanced distribution of gain and loss, where $\Re{\epsilon(r)} = \Re{\epsilon(-r)}$ and $\Im{\epsilon(r)} = -\Im{\epsilon(-r)}$ (Fig.~\ref{fig:1}(a)) \cite{Feng2017,ozdemir2019parity}.
Under this condition, frequency gaps of real eigenvalues can still be obtained (Fig.~\ref{fig:1}(b)) in the $\mathcal{PT}$-symmetric phase despite the presence of non-Hermiticity.

More recently, time-modulated photonic systems, represented by photonic time-modulated crystals (PTCs), have attracted much attention as they provide a promising platform for manipulating photonic responses through temporal variation of material optical properties such as permittivity~\cite{galiffi2022photonics, Lustig:23, lustig2018topological,hayran2022homega, pacheco2022time,asgari2024theory}. 
This dynamic modulation and the resulting PTCs support the engineering of photonic Floquet band structures through temporal control and facilitate the manipulation of Floquet phases, time-domain reflection and diffraction, amplified emission and lasing, harmonic generation, heat transfer, and quantum light generation~\cite{lyubarov2022amplified,moussa2023observation,tirole2023double,zhou2020broadband,lustig2023time,wang2023metasurface,horsley2023quantum,galiffi2019broadband,franke2021fermi,tirole2024second,park2022revealing,yu2024time,park2024spontaneous,bae2025cavity,sustaeta2025quantum}.
In contrast to the frequency gap in spatial crystals, a striking feature of these time-modulated photonic systems is the appearance of momentum gaps~\cite{lustig2018topological}. 
Considering these rapid advances, it is pertinent to investigate whether there is a temporal counterpart to the \(\mathcal{PT}\)-symmetry widely explored in the spatial domain.
Li et.~al.~\cite{li2021temporal} introduced this type of temporal $\mathcal{PT}$-symmetry to photonic temporal interfaces, where extreme energy manipulation is achieved in the non-Hermitian regime. However, the Hermitian regime and associated $\mathcal{PT}$-symmetry phase transition, as anticipated from their spatial duals, remain elusive in the temporal domain.

In this work, we theoretically uncover the existence of $\pt$-symmetry phase transition in photonic time-modulated media. 
To this end, we extend the transfer matrix method (TMM) for photonic time-modulated media from binary-level to multi-level scenarios and further derive a continuous differential framework that simultaneously accommodates modulations in both the real and imaginary components of the permittivity $\epsilon(t)$.
Using Magnus expansion analysis, we rigorously demonstrate that the $\pt$-symmetric phase persists under single-frequency modulation when the modulations of $\Re{\epsilon(t)}$ and $\Im{\epsilon(t)}$ exhibit a $\mathrm{\pi}/2$ phase delay and the modulation depth of $\Im{\epsilon(t)}$ exceeds that of $\Re{\epsilon(t)}$. Violation of these conditions leads to the emergence of the $\pt$-broken phase.
This \(\mathcal{PT}\) symmetry phase transition is supported by our full-wave simulations and analytical calculation of pulse scattering in time-modulated media in both the Hermitian and non-Hermitian regimes.
%


A time-domain version of the $\mathcal{PT}$-symmetry in a homogeneous but time-varying photonic medium of permittivity $\epsilon(t)$ could be defined as 
$
\epsilon(-t) = \epsilon^*(t),
$
where the asterisk denotes complex conjugation.
More specifically, for a periodic modulation of $\epsilon(t)$, in each period
there exists a time axis (chosen as $t=0$ in Fig.~\ref{fig:1} without loss of generality) such that the real part of the permittivity is symmetric in time, $\Re{\epsilon(t)} = \Re{\epsilon(-t)}$, and the imaginary part is antisymmetric, $\Im{\epsilon(t)} = -\Im{\epsilon(-t)}$.
In quantum mechanics, this condition in a homogeneous potential naturally leads to real spectra (see Supplemental Material (SM S1)), whereas the situation becomes complicated for photonic media.

To discuss the consequences of this temporal $\mathcal{PT}$-symmetry in photonics, we extend the transfer matrix method to handle modulation in both $\Re{\epsilon(t)}$ and $\Im{\epsilon(t)}$ from binary-level modulation to multi-level and continuous modulation schemes.
Starting with binary-state modulation, we consider systems where the permittivity alternates between two discrete values periodically in time. This modulation can be characterized using transfer matrices that describe the continuity of electromagnetic fields at temporal boundaries, which leads to the dispersion relation~\cite{lustig2018topological}:
\begin{align}
    \Omega(k_z)T = \cos^{-1}\left[\text{Tr}(M)/2\right],
    \label{eq:TMM_2state}
\end{align}
where $\Omega$ is the quasienergy, $k_z$ is the wavevector along the wave's propagation direction.
We generalize this approach to $n$-state modulation schemes, where the permittivity cycles through $n$ discrete values within each period. The total transfer matrix $M$ over one period is constructed as the product of matrices corresponding to each temporal segment:
\begin{widetext}
    \begin{equation}
    \label{eq:multistate}
    \begin{split}
    M &= \frac12
    \begin{pmatrix}
      f_{1} & -f_{1}/p_{1} \\
      1/f_{1} & 1/(p_{1}f_{1})
    \end{pmatrix}
    \underbrace{\left[
    \prod_{k=1}^{n-1}
    \begin{pmatrix}
      1/f_{2k} & f_{2k} \\
      -1/f_{2k} & f_{2k}
    \end{pmatrix}
    \;\frac12\;
    \begin{pmatrix}
      f_{2k+1} & -f_{2k+1}/p_{k+1} \\
      1/f_{2k+1} & 1/(p_{k+1}f_{2k+1})
    \end{pmatrix}
    \right]}_{U}
    \begin{pmatrix}
      1/f_{2n} & f_{2n} \\
      -1/f_{2n} & f_{2n}
    \end{pmatrix}
    \end{split},
    \end{equation}
\end{widetext}
where \(f_k = \exp(\iu\omega_k\tau_k)\), \(\tau_k = \sum_{i=1}^{k}t_{\mathrm{i}}\), \(\omega_k\) is the angular frequency of the \(k\)-th time segment, \(p_k = \omega_k/\omega_{k+1}\) for \(k=1,\ldots,n-1\), and \(p_n = \omega_n/\omega_1\)(SM S2).

We explicitly prove the realness of the trace of the transfer matrix $M$ under the temporal $\mathcal{PT}$-symmetry (see details in SM S3). In short, the $\pt$-symmetry imposes
\begin{align}
M^{\dagger} = \mathcal{P} M \mathcal{P}^{-1},\qquad 
\label{eq:PT}
\end{align}
where we find $\mathcal{P}=\sigma_z$. Eq.~\eqref{eq:PT} is known as the pseudo-Hermiticity condition under operator $\mathcal{P}$~\cite{10.1063/1.1418246,PhysRevLett.89.270401}. 
Eq.~\eqref{eq:PT} ensures $\operatorname{Tr} M = \operatorname{Tr} M^{\dagger} \in \mathbb{R}$ and constrains the eigenvalues of $M$ to either being a degenerate real pair or forming a complex-conjugate pair.

If the permittivity $\epsilon(t)$ varies continuously with time, we can further reformulate the discrete matrix product Eq.~\eqref{eq:multistate} into a continuous differential framework. By considering infinitesimal time intervals, we derive a first-order matrix differential equation (SM S2):
\begin{align}
\frac{\mathrm{d}U(t)}{\mathrm{d}t} = U(t) R(t).
\end{align}
Here $U(t)$ is a continuous formulation of the intermediate $2n-2$ matrices in Eq.~\eqref{eq:multistate}, and 
$R(t)=\begin{pmatrix}
0 & \iu\omega(t)\\
\iu\omega(t) & \omega'(t)/\omega(t)
\end{pmatrix}$ 
is the instantaneous generator matrix,
where $ \omega(t;k_z)=ck_z/\sqrt{\epsilon(t)}$.
In the following context, we denote the full-cycle \(U(T)\) implicitly as \(U\) for convenience, while \(U(t)\) refers to the general time-dependent evolution matrix.
The dispersion relation can be written as 
\begin{align}
    \Omega(k_z)T =  \cos^{-1}\left[\text{Tr}(U)/2\right],
\label{eq:eigen}
\end{align}
which remains consistent with Eq.~\eqref{eq:TMM_2state} because in the limit \(n \to \infty\), the rightmost and leftmost matrices in Eq.~\eqref{eq:multistate} act as a similarity transformation on the intermediate matrix, and the trace remains invariant under this transformation. The numerical consistency of this convergence is demonstrated in SM S4.

Therefore, to ensure that the Floquet band structure remains Hermitian, it is required that $\text{Tr}(U) = u_{11} + u_{22}$ is real and satisfies $|\text{Tr}(U)| \leq 2$, where $u_{11}$ and $u_{22}$ are the diagonal elements of $U$.
We note that this framework can even handle discontinuous modulation where a mathematical renormalization of the $u_{22}$ component can be introduced (see SM S2).

We can further define differential operators to encapsulate the behavior of the system:
\begin{align}
\hat{O} &= \left( \dfrac{\mathrm{d}}{\mathrm{d}t} \right)^2 + \dfrac{(c k_z)^2}{\epsilon(t)} + \dfrac{1}{2} \dfrac{\epsilon'(t)}{\epsilon(t)} \dfrac{\mathrm{d}}{\mathrm{d}t}, \\
\hat{P} &= \dfrac{1}{2} \dfrac{\epsilon'(t)}{\epsilon(t)} \dfrac{\mathrm{d}}{\mathrm{d}t}.
\end{align}
The $U(t)$ matrix should satisfy:
\begin{align}
(\hat{O} + \hat{P}) u_{11}(t) = 0, \\
(\hat{O} - \hat{P}) {u}_{22}(t) = 0.
\end{align}
This mathematical formulation will be helpful for our further analysis on the $\pt$ phase transition.
One may equivalently derive the differential operators from Maxwell's equations, where the $u_{11}$ and $u_{22}$ parameters correspond to the normalized magnetic and displacement fields, respectively (see SM S2). 
Crucially, the operator $\hat{O}$ satisfies:
\begin{align}
\hat{O} \cos\left( \int_{0}^{t} \omega(\tau; k_z) \, \mathrm{d}\tau \right) = 0,
\end{align}
regardless of the specific form of the modulation $\epsilon(t)$; see SM S5. 

An example satisfying the requirement of the temporal $\pt$-symmetry is the sinusoidal modulation of $\epsilon(t)$, where the real and imaginary components share the same frequency but are phase-shifted by $\mathrm{\pi}/2$.
Specifically, one can express 
\begin{align}
\epsilon(t) = \epsilon_{\mathrm{i}} + \epsilon_A \left[ \cos\left( \omega_\text{m} t \right) + \iu \gamma \sin\left(\omega_\text{m} t \right) \right],
\label{eq:single_freq}
\end{align}
where \( \omega_\text{m} = 2\mathrm{\pi}/T \) is the fundamental angular frequency, $\epsilon_{\mathrm{i}} > \epsilon_A$ to ensure that $\Re\epsilon(t)$ remains positive. The contrast between the modulation depth of $\Im\epsilon(t)$ to that of $\Re\epsilon(t)$ is defined as $\gamma$.
\begin{figure}[h!]
    \centering
    \includegraphics[width=1\linewidth]{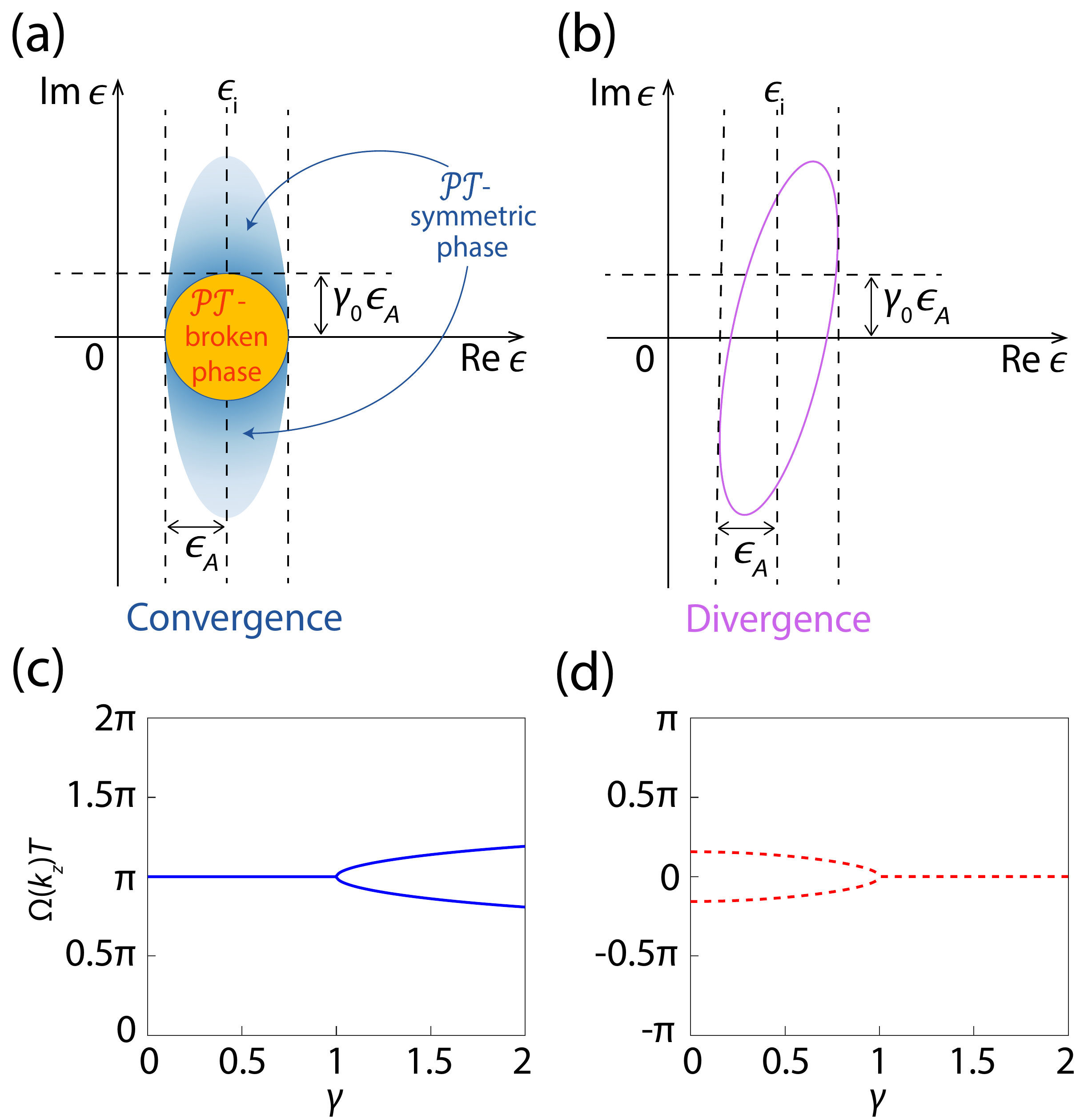} 
    \caption{\textbf{Temporal \(\mathcal{PT}\)-symmetric and \(\mathcal{PT}\)-broken behavior under sinusoidal modulation of \(\epsilon(t)\).}
    (a-b) Complex-plane trajectories of \(\epsilon(t)\) under sinusoidal modulation. In the convergent case (a), \(\Re{\epsilon(t)}\) and \(\Im{\epsilon(t)}\) oscillate at the same frequency with a phase difference of \(\mathrm{\pi}/2\), forming \(\mathcal{PT}\)-symmetric Hermitian (blue ellipse) and \(\mathcal{PT}\)-broken non-Hermitian (orange ellipse) regions. This behavior corresponds to convergent bounded oscillations of \(\text{Tr}(U)\). Divergence of \(\text{Tr}(U)\) appears if the modulation phase difference deviates from $\pm\mathrm{\pi}/2$ (b). 
    (c-d) \(\mathcal{PT}\)-symmetric and \(\mathcal{PT}\)-broken phase transitions with varying modulation depth contrast \(\gamma\): the real (c) and imaginary (d) parts of \(\Omega(k_z)T\) at the first extremum of \(\text{Tr}(U)/2-k_z\) are plotted with parameters \(\epsilon_{\mathrm{i}} = 1.8\) and \(\epsilon_A = 1\).
}
    \label{fig:5}
\end{figure}

As illustrated in Fig.~\ref{fig:5}(a), the trajectory of $\epsilon(t)$ in the complex plane can be plotted in polar coordinates over one complete evolution period. With an exact phase difference of $\pm\pi/2$ between $\Re\epsilon$ and $\Im\epsilon$, the trajectory exhibits mirror symmetry about the real axis $\Re\epsilon$, which satisfies temporal $\mathcal{PT}$-symmetry. In this case, $\Tr(U)$ remains purely real for all $k_z$, and the corresponding $\Omega(k_z)$ is convergent.
Conversely, if the phase difference between the real and imaginary parts deviates from $\pm\pi/2$ (Fig.~\ref{fig:5}(b)), the trajectory is no longer mirror-symmetric about $\Re\epsilon$, thus breaking temporal $\mathcal{PT}$-symmetry. As a result, $\Tr(U)$ acquires a nonzero imaginary part for generic $k_z$, and $\Omega(k_z)$ becomes divergent.

The Hermiticity of the band $\Omega(k_z)$ requires both $\Tr(U)\in\mathbb{R}$ and $\abs{\Tr(U)}\leq2$. As we have proved previously that a $\mathcal{PT}$-symmetric system automatically satisfy the first criterion, the Hermiticity reduces to the condition $\abs{\Tr(U)}\le2$, which hinges critically on the modulation depth contrast $\gamma$ for the single-frequency modulation in Eq.~\eqref{eq:single_freq}.

As shown in Fig.~\ref{fig:5}(c) and (d), $\gamma$ is the key parameter affecting the $\pt$-symmetry transition, playing a similar role as the spatial gain-loss contrast for the spatial $\pt$-transition~\cite{Feng2017}.
For $\gamma < \gamma_0$ (yellow circular region in Fig.~\ref{fig:5}), the system exhibits non-Hermitian energy bands.
For $\gamma > \gamma_0$ (blue region in Fig.~\ref{fig:5}), the system becomes Hermitian.
$\gamma = \gamma_0$ corresponds to the exceptional point where the temporal $\pt$ phase transition occurs.

For the single frequency modulation in Eq.~\eqref{eq:single_freq}, the phase transition occurs at $\gamma_0 \approx 1$, which will be explained by our Magnus expansion analysis (see SM S6).
Interestingly, in contrast to the spatial case where the $\pt$-symmetric phase appears for a weak loss-gain contrast, in the temporal case, the $\pt$-symmetric phase requires stronger $\Im\epsilon$ modulation than $\Re\epsilon$ modulation.

We provide more analytical insights to explain the $\pt$ phase transition in Fig.~\ref{fig:5}(c,d) using the Magnus expansion~\cite{https://doi.org/10.1002/cpa.3160070404}.
We adopt a perturbative analysis by treating \(\hat{P}=\frac{1}{2}\epsilon'(t)/\epsilon(t)\,\partial_t\) as a perturbation to $\hat{O}$. We therefore write
\(u_{11}=u_{11}^{(0)}+u_{11}^{(1)}+u_{11}^{(2)}+\dots\), \(u_{22}=u_{22}^{(0)}+u_{22}^{(1)}+u_{22}^{(2)}+\dots\),
and introduce the summation and difference combinations
\(S(t)=u_{11}(t)+u_{22}(t)\), \(D(t)=u_{11}(t)-u_{22}(t)\).
Adding and subtracting the evolution equations \((\hat{O}\pm\hat{P})u_{11,22}=0\) gives the coupled non-Hermitian system
\(\hat{O}S(t)=-\hat{P}D(t)\) and \(\hat{O}D(t)=-\hat{P}S(t)\).

At zeroth order, the unperturbed solution of \(\hat{O}S_{0}=0\) is \(S_{0}(t)=u_{11}^{(0)}+u_{22}^{(0)}=2\cos\theta(t)\), with accumulated phase \(\theta(t)=\int_{0}^{t}\omega(\tau)\,\mathrm{d}\tau\).  Hence, the extremum value $\lvert\text{Tr}(U)\rvert = 2$, regardless of $\gamma$, shows that the zeroth-order approximation fails to capture any phase transition.
At first order, from \(\hat{O}D_{1}=-\hat{P}S_{0}\) we obtain  
\(D_{1}(t)=\int_{0}^{t} \sin\bigl[\theta(t)-\theta(t_{1})\bigr] L(t_{1})\sin\theta(t_{1})\,\mathrm{d}t_{1}\), \(L(t)=\epsilon'(t)/\epsilon(t)\), whereas \(S_{1}(t)=0\).
At second order, with \(\hat{O}S_{2}=-\hat{P}D_{1}\) the symmetric component becomes  
\begin{align}
S_{2}(t)  
&=-\frac{1}{2}\int_{0}^{t}\!\mathrm{d}t_{1}\,L(t_{1})  
\sin\bigl[\theta(t)-\theta(t_{1})\bigr]  
\nonumber\\  
&\phantom{={}}\times  
\int_{0}^{t_{1}}\!\mathrm{d}t_{2}\,L(t_{2})  
\sin\theta(t_{2})  
\cos\bigl[\theta(t_{1})-\theta(t_{2})\bigr],  
\label{eq:S2}  
\end{align}
so that \(u_{11}+u_{22}=2\cos\theta+S_{2}+\mathcal{O}(\hat{P}^{3})\). Although the zeroth-order approximation yields the universal trivial extrema $\lvert\text{Tr}(U)\rvert = 2$, the addition of $S_{2}$ lifts this degeneracy, makes the system sensitive to $\gamma$, and thus captures the onset of phase transitions.

\begin{figure}[h!]
    \centering
    \includegraphics[width=1\linewidth]{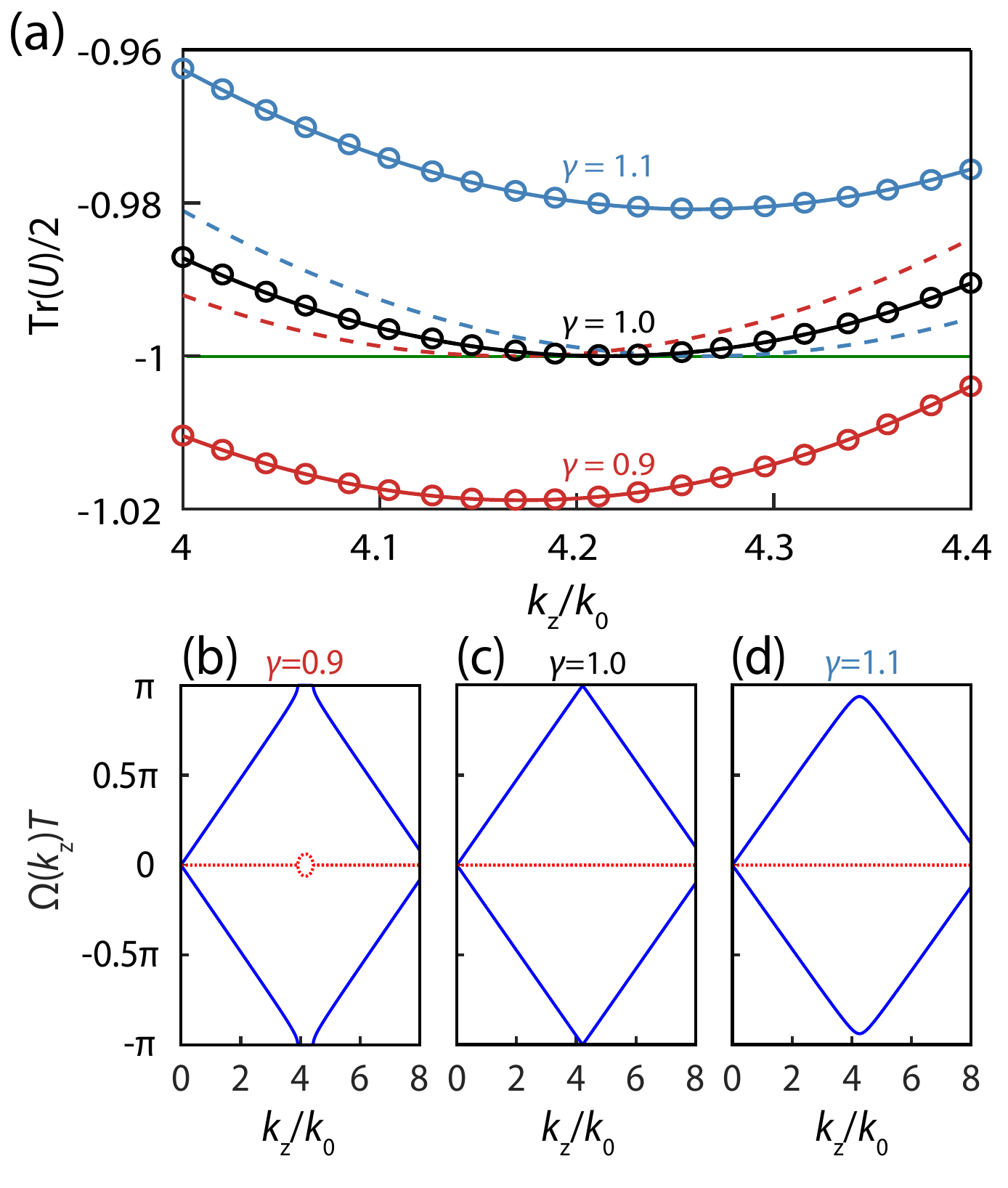}
    \caption{
    \textbf{Magnus expansion analysis of the $\mathcal{PT}$ phase transition and band structures.} (a) Open circles, dashed lines, and solid lines are direct numerical ODE solutions, zeroth-order, and second-order Magnus analysis results, respectively. Fixed parameters: $\epsilon_{\mathrm{i}}=1.8$, $k_0 = \omega_\text{m} / (2\mathrm{\pi} c)$. The horizontal green line indicates $\mathrm{Tr}(U)/2 = -1$. (b-d) Blue solid lines show the real parts of the $\Omega(k_z)T$, while red dashed lines represent the imaginary parts, for three different values of $\gamma=\{0.9, 1.0, 1.1\}$. 
}
    \label{fig:mg}
\end{figure}

For the specific permittivity modulation given in Eq.\,\eqref{eq:single_freq} with $\epsilon_A=1$ without loss of generality and $\epsilon_{\mathrm{i}}>1$, the phase accumulation over one period can be compactly written as the contour integral.
$
 \theta(T) = c\,k_{z}/(\iu\,\omega_\text{m}) \oint_{|z|=1} \mathrm{d}z / [ z^{3/2} \sqrt{ \epsilon_{\mathrm{i}}\,z + (1/2)(1-\gamma) + (1/2)(1+\gamma)z^{2} } ] 
$.  
$\theta(T)$ reduces to \(2\mathrm{\pi}\,c\,k_{z}/(\omega_\text{m}\sqrt{\epsilon_{\mathrm{i}}})\) when \(\gamma=1\). The function \(L(t)\) takes the explicit form:  
$
L(t) = \omega_\text{m} \bigl[ -\sin(\omega_\text{m} t) + \iu\gamma \cos(\omega_\text{m} t) \bigr] 
/ \bigl[ \epsilon_{\mathrm{i}} + \cos(\omega_\text{m} t) + \iu\gamma \sin(\omega_\text{m} t) \bigr].
$
Substituting \(L(t)\) into the second-order expression Eq.~\eqref{eq:S2} yields the trace expansion:  
\begin{widetext}  
\begin{align}  
\Tr(U)  
=2\cos\theta(T) &- \frac{1}{2}\int_{0}^{T}\!\mathrm{d}t_{1}  
\frac{\omega_\text{m}\bigl[-\sin(\omega_\text{m}t_{1})+\iu\gamma\cos(\omega_\text{m}t_{1})\bigr]}{\epsilon_{\mathrm{i}}+\cos(\omega_\text{m}t_{1})+\iu\gamma\sin(\omega_\text{m}t_{1})}  
\sin\bigl[\theta(T)-\theta(t_{1})\bigr] \notag \\  
&\quad \times \int_{0}^{t_{1}}\!\mathrm{d}t_{2}  
\frac{\omega_\text{m}\bigl[-\sin(\omega_\text{m}t_{2})+\iu\gamma\cos(\omega_\text{m}t_{2})\bigr]}{\epsilon_{\mathrm{i}}+\cos(\omega_\text{m}t_{2})+\iu\gamma\sin(\omega_\text{m}t_{2})}  
\sin\theta(t_{2})  
\cos\bigl[\theta(t_{1})-\theta(t_{2})\bigr]  
+\mathcal{O}(L^{3}),  
\end{align}  
\end{widetext}  
with \(\theta(t)=\int_{0}^{t} \left[ ck_{z}\,\mathrm{d}\tau \big/ \sqrt{ \epsilon_{\mathrm{i}}+\cos(\omega_\text{m}\tau)+\iu\gamma\sin(\omega_\text{m}\tau) } \right]
\).
As shown in Fig.~\ref{fig:mg}(a), even at zeroth order, the Magnus expansion already captures the fundamental oscillation frequency observed in direct numerical integration of the ODE; for sufficiently large \(k_z\), the zeroth-order result agrees very well with the full solution for selected values of \(\gamma\). In the special case \(\gamma = 1\), this agreement persists across the entire \(k_z\) domain, reflecting the fact that \(S_2(T)=0\) under this condition (see SM S6), so the zeroth- and second-order expansions are identical. 
When carried out to the second order, the Magnus approximation exhibits excellent agreement with the direct ODE solutions over the full range of \(k_z\) for all choices of~$\gamma$.
Their corresponding band structures indeed exhibit the predicted non-Hermitian to Hermitian $\pt$ phase transition as $\gamma$ increases (Fig.~\ref{fig:mg}(b-d)).

\begin{figure*}[htbp]
    \centering
    \includegraphics[width=\linewidth]{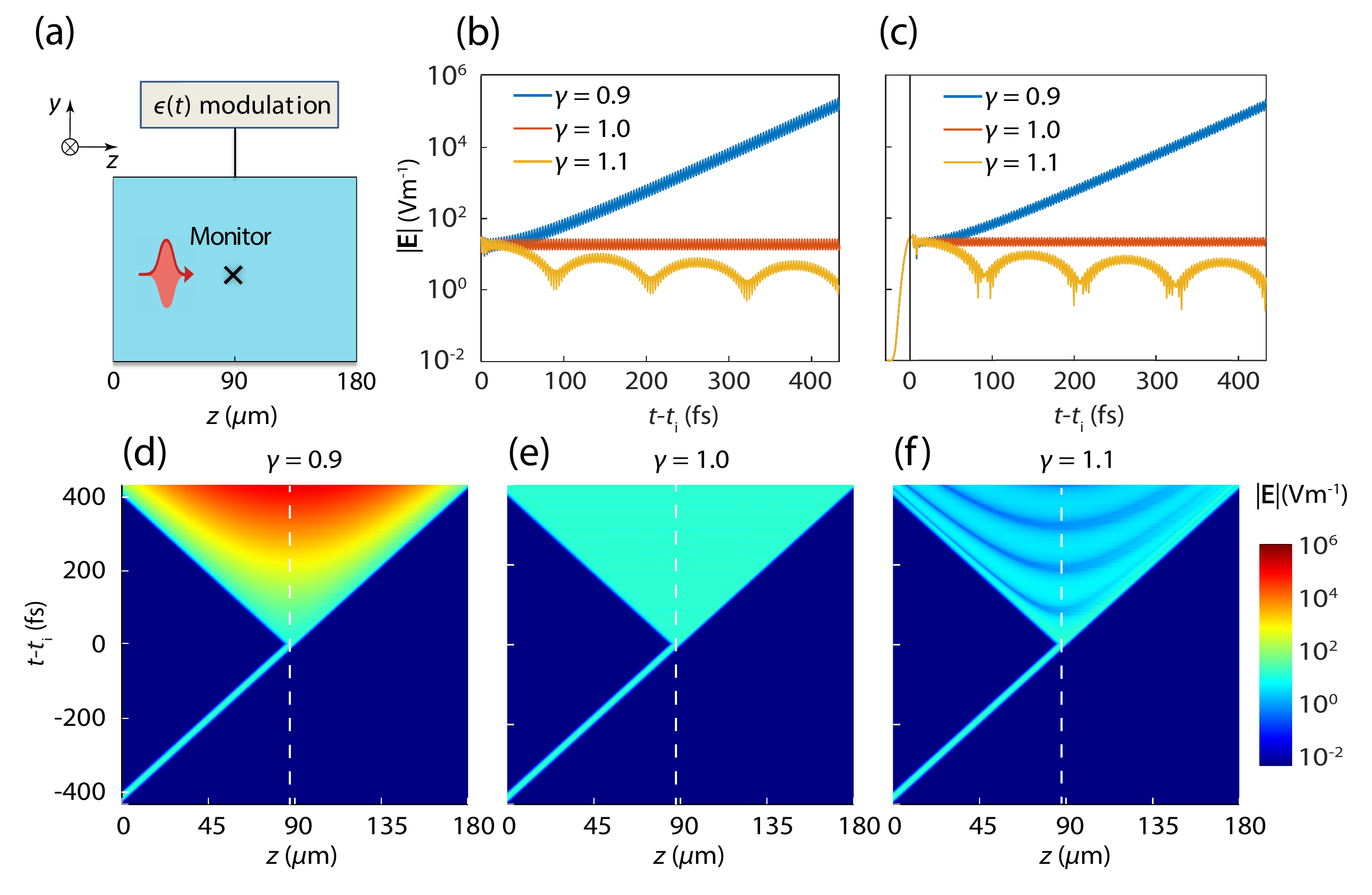}
    \caption{%
    \textbf{Validation of the appearance of the temporal $\mathcal{PT}$-symmetric transition.
    }
(a) Schematic of the simulation setup, showing the propagation of an electromagnetic pulse in a dielectric medium with time-dependent permittivity $\epsilon(t)$. The monitor at $z_0=\SI{87.3}{\um}$ records the waveform as illustrated in (c).
(b) Analytically calculated amplitude of $|\mathbf{E}(z,t)|$ from the band structure for $\gamma = 0.9$ (blue), $\gamma = 1.0$ (red), and $\gamma = 1.1$ (yellow). Each rapid oscillation corresponds to one modulation period, and the amplitude envelope exhibits exponential growth when $\gamma = 0.9$, stationarity when $\gamma = 1.0$, and non-monotonic decay when $\gamma = 1.1$.
(c) Numerically calculated amplitude of $|\mathbf{E}(z_0,t)|$. The modulation is activated at $ t_{\mathrm{i}}=\SI{433.63}{\fs}$.
(d--f) Logarithmic spatiotemporal evolution of $|\mathbf{E}(z,t)|$ from simulations for $\gamma = 0.9$ (d), $\gamma = 1.0$ (e), and $\gamma = 1.1$ (f), respectively. The vertical dashed line cuts at $z_0$ correspond to the waveform shown in (c).
}
    \label{fig:FDTD}
\end{figure*}

We confirm the existence of the $\mathcal{PT}$-transition with dynamics calculated from both analytical and numerical results.
Parameters and the detailed configuration of the numerical setup are described in SM S7.
The ratio $\gamma$ is varied in the set $\{0.9, 1.0, 1.1\}$, which correspond, respectively, to the $\mathcal{PT}$-broken, $\mathcal{PT}$-transition, and $\mathcal{PT}$-symmetric phases.

We can analytically predict the dynamics using the calculated Floquet band structure. For Fig.~\ref{fig:FDTD}(b) we write the initial pulse as a Gaussian superposition of plane waves in the displacement field, $D(z,t_{\mathrm{i}})=\int A(k_z)\exp(\iu k_zz)\,\mathrm{d}k_z$ with $A(k_z)\propto\exp[-(k_z-k_c)^2/2\sigma_k^2]$. 
At a fixed position each $k_z$ component evolves as $D(k_z,t)=D(k_z,t_{\mathrm{i}})\exp[-\iu\Omega(k_z)(t-t_{\mathrm{i}})]$, where $\Omega(k_z)$ is the Floquet exponent extracted from the period-$T$ monodromy matrix associated with the modulated permittivity $\epsilon_r(t)$.  
Integrating over $k_z$ gives $D(z=0,t) = \int D(k_z,t)\,\mathrm{d}k_z\,.$
Electric fields $E(z=0,t)=D(z=0,t)/[\epsilon_0\epsilon_r(t)]$ at the monitor position are plotted for $\gamma = \{0.9,\, 1.0,\, 1.1\}$ in Fig.~\ref{fig:FDTD}(b), respectively.

The predicted dynamics is directly validated by the full-wave finite-element numerical simulations (Fig.~\ref{fig:FDTD}(c-f)). 
The entire spatiotemporal behaviors are shown in Fig.~\ref{fig:FDTD}(d-f) and the cut (white dashed line) at the monitor position is plotted in Fig.~\ref{fig:FDTD}(c), which agrees well with the analytical calculation in Fig.~\ref{fig:FDTD}(b).
The minor difference between Fig.~\ref{fig:FDTD}(b) and Fig.~\ref{fig:FDTD}(c) stems from the fact that treating $\Omega(k_z)$ as time-independent captures the correct discrete values at each period-end but neglects subtle variations within each period.

When $\gamma=0.9$, the waveform exhibits exponential growth since the excitation frequency is within the non-Hermitian gap (Fig.~\ref{fig:mg}b). When $\gamma = 1.0$, near the EP transition in the $\gamma$ parameter space, the amplitude remains constant after each period, agreeing with the nearly linear Floquet dispersion (Fig.~\ref{fig:mg}c). In contrast, when $\gamma = 1.1$, a slowly decaying, non-monotonic envelope appears due to the differences in the group velocity and phase mismatch of the Hermitian band (Fig.~\ref{fig:mg}d), resulting in the alternating appearances of field maxima and minima.
Such evolution between exponential growth and oscillatory decay is further confirmed by the spatiotemporal field distribution in Figures~\ref{fig:FDTD}(d--f).

We prove that in the context of complex permittivity modulation, $\pt$-symmetry is a necessary condition of real spectra, and demonstrate the existence of a parity-time (PT) symmetry phase transition in photonic time-modulated media.
In the sinusoidal modulation of the real and imaginary permittivity components with a critical $\pi/2$ phase difference, the modulation depth governs the transition between Hermitian and non-Hermitian regimes. 
By extending the transfer matrix method to a continuous differential framework and employing Magnus expansion analysis, we establish that the $\pt$-symmetric phase emerges when the modulation depth of the imaginary permittivity exceeds that of the real component ($\gamma > \gamma_0 \approx 1$). Full-wave simulations and analytical pulse dynamics confirm distinct regimes: exponential amplification in the $\pt$-broken phase, stationary amplitude at the exceptional point, and oscillatory decay in the $\pt$-symmetric phase.

The experimental realization of the predicted $\pt$-phase transition could leverage platforms that enable synchronized permittivity modulation, such as optically pumped epsilon-near-zero materials, active metasurfaces and circuits integrated with dynamic elements---such as varactor diodes, tunable capacitors, and resistors for the microwave regimes.
Further integration of this temporal $\pt$-symmetry with its spatial counterpart could pave the way for spatiotemporal synthetic photonic materials with enhanced control over light-matter interactions.

\begin{acknowledgement}

The author thanks Bengy Tsz Tsun Wong, Yihao Yang, and Shuang Zhang for helpful discussions.

\end{acknowledgement}


%

\end{document}